\documentclass[aps,prl,twocolumn,amsmath,amssymb,nofootinbib,superscriptaddress,floatfix]{revtex4}

\usepackage{amsmath}
\usepackage{amssymb}
\usepackage{amsthm}
\usepackage{graphicx}
\usepackage{ulem}
\usepackage{bm}
\usepackage[hypertex]{hyperref}
\usepackage{subfigure}

\begin{document}

\title{Surface Code Threshold in the Presence of Correlated
  Errors}

\author{E. Novais}

\affiliation{Centro de Ci\^encias Naturais e Humanas, Universidade
  Federal do ABC, Santo Andr\'e, SP, Brazil}

\author{Eduardo R. Mucciolo}

\affiliation{Department of Physics, University of Central Florida,
  Orlando, Florida 32816, USA}

\date{\today}

\begin{abstract}
We study the fidelity of the surface code in the presence of
correlated errors induced by the coupling of physical qubits to a
bosonic environment. By mapping the time evolution of the system after
one quantum error correction cycle onto a statistical spin model, we
show that the existence of an error threshold is related to the
appearance of an order-disorder phase transition in the statistical
model in the thermodynamic limit. This allows us to relate the error
threshold to bath parameters and to the spatial range of the
correlated errors.
\end{abstract}

\maketitle


The surface code is considered one of the best quantum error
correction (QEC) codes to implement on physical devices
\cite{DKP2002,DiVincenzo09-a,YWC+12}. This stems from two major
points: First, all syndromes and operations can be performed with
spatially local operators; second, all threshold estimates show that,
for sufficiently large lattices, the error threshold is the highest
known for two-dimensional architectures with only nearest-neighbor
interactions \cite{RH07,FWH11,bombin12}.

The error threshold is usually defined for stochastic error models. By
assuming that errors are independent events and assigning a
probability $p$ to each of these events, it has been shown numerically
that the quantum information encoded in the surface can be faithfully
protected when $p$ is below a critical value. Although this result is
firmly established numerically, there are two big open questions that
still need to be addressed. First, stochastic error models are
approximations to reality that sometimes cannot be justified. In fact,
most studies so far lacked a microscopic description of the
interaction between physical qubits and the environment. Second, the
same locality of operations and syndromes that makes the surface code
powerful also makes it more susceptible to correlated errors. Thus, a
discussion of the tradeoff between locality of operations and
correlated errors is long overdue. In this Letter, we address both
issues by employing a more realistic error model. We consider a
Caldeira-Leggett type of environment where freely propagating bosonic
modes couple linearly and locally to the physical qubits. Such a model
has a very strong physical motivation since in most experimental
implementations photons and phonons couple to the two-level systems
making up the physical qubits. This model also plays a fundamental
role in our understanding of decoherence \cite{Unruh:1995:992} and its
interplay with QEC \cite{Novais:2006:040501, Novais:2007:040501,
  Novais2008, Ng:0810.4953v1, NERM10,Preskill.2012}.

Consider a logical qubit in a quantum memory. In a QEC cycle, the
logical qubit is prepared and left to freely evolve during a certain
time interval. Then, syndromes are extracted, and, if necessary,
suitable error correction operations are implemented to bring the
logical qubit back to its original state. In our analysis, we evaluate
the fidelity of a logical qubit after such a QEC cycle under the
assumption of nonerror syndromes. This assumption is not essential to
our results, but makes the calculation more concise. In addition, we
assume that the bath is initially at zero temperature and that it is
reset to this temperature at the end of the QEC cycle.

Our results show a sharp transition between two distinct noise
regimes. On one hand, below a fictitious critical temperature (which
is related to microscopic parameters of the bath), the dissipation due
to the bath cannot be suppressed by the encoding. On the other hand,
above this critical temperature, we show that if the thermodynamic
limit is taken, the effects of the bosonic environment become
irrelevant and the logical qubit is fully protected.

Even though this investigation focuses on quantum information
protection, the physical problem we consider has a much broader
appeal. In essence, it amounts to a lattice gauge system interacting
with a scalar bosonic field in two dimensions \cite{Kogut}. In this
language, the nonexistence of a quantum error threshold can be
understood as the lifting of the ground state topological degeneracy.
Our discussion of the error threshold can therefore be recast as a
quantum phase transition. This fits into our earlier
discussion of the threshold theorem as resembling a quantum phase
transition \cite{Novais:2007:040501}. This analogy is nontrivial since
the error threshold is in essence a driven dynamical problem, far from
the equilibrium conditions required to describe phase transitions in
statistical mechanics. This work turns what was an analogy into a
well-defined map. Therefore, we believe that the results presented
here transcend our original motivation and complement the recent
discussion in Ref. \onlinecite{Preskill.2012}.

\begin{figure}
\centering \subfigure{
  \includegraphics[width=3.5cm]{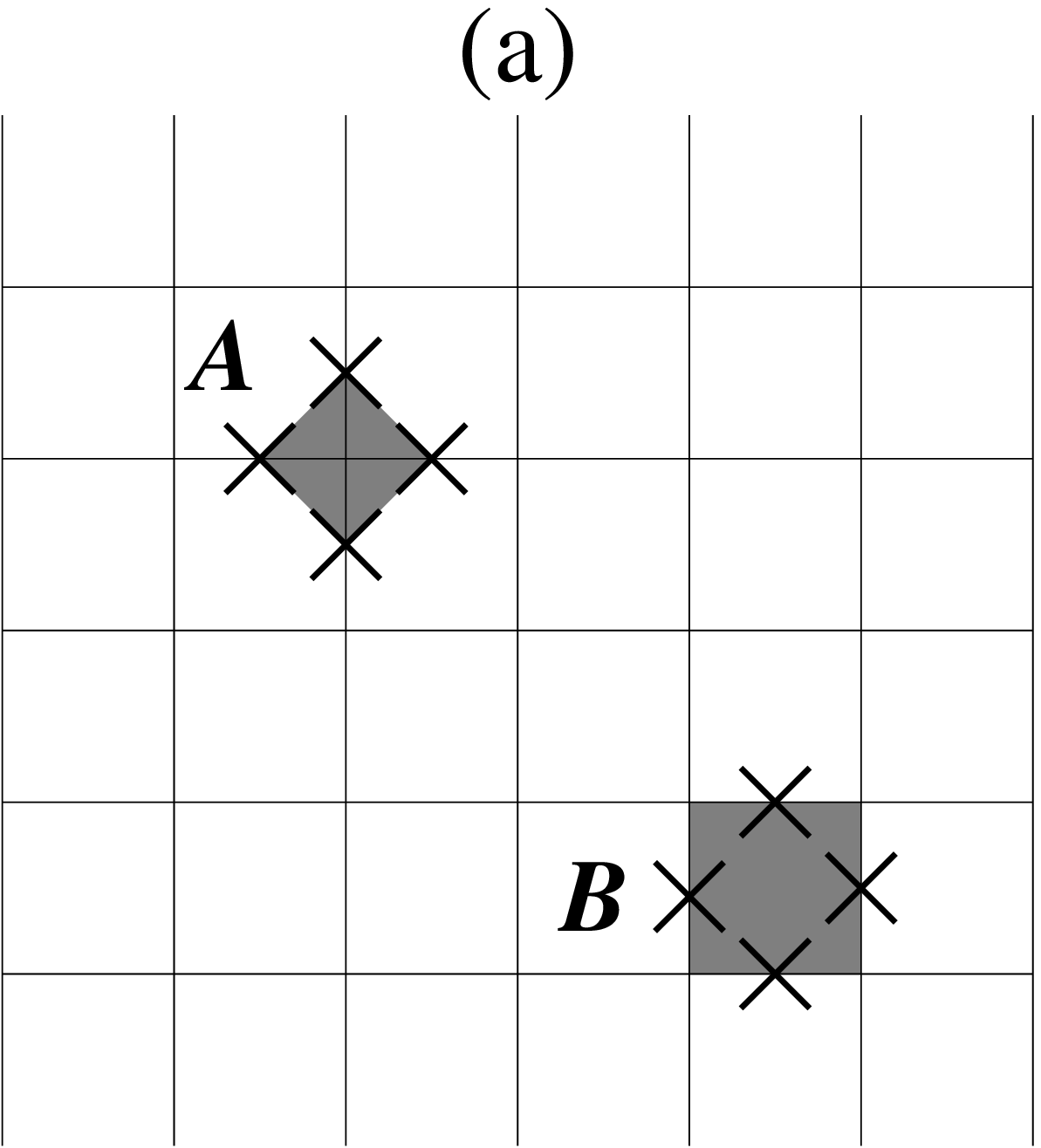}}
\hspace{.5cm}
 \subfigure{
  \includegraphics[width=3.85cm]{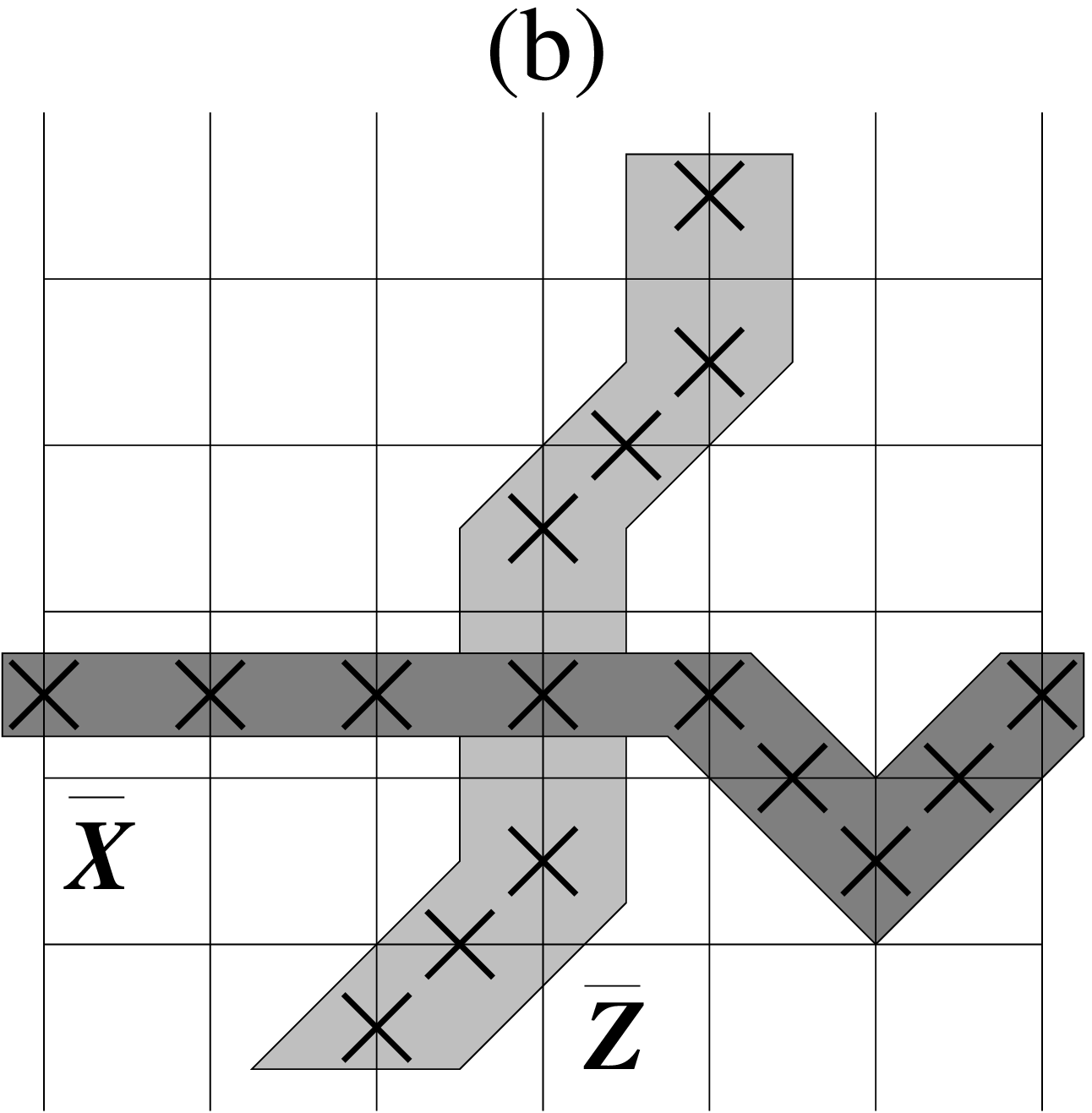}}
\caption{(a) Star ($A$) and plaquette ($B$) operators. (b) Examples of
  logical $\bar{X}$ and $\bar{Z}$ operators. The crosses and shades
  indicate the qubits involved in the respective operator.}
\label{fig:encoding}
\end{figure}

{\it The model} --We consider physical qubits
$\{\vec{\sigma}_i\}_{i=1,\ldots,N}$ (i.e., spin 1/2 systems) located
on the links of a square lattice 
with open boundary conditions (see Fig. \ref{fig:encoding}).
The QEC code is defined by an encoding prescription and a set of
stabilizer operators \cite{Gottesman}. The stabilizer operators of the
surface code are easily labeled when we define stars and plaquette
operators. Star operators are the product of the four $\sigma^x$
operators of qubits adjacent to a vertex of the square lattice,
$A_{\Diamond} = \prod_{i\in\diamondsuit} \sigma_{i}^{x}$. Similarly,
plaquette operators are the product of four $\sigma^z$ of qubits
located at the edges of a tile of the square lattice, $B_{\square} =
\prod_{i\in\square} \sigma_{i}^{z}$. Each plaquette and star is a
stabilizer that has to be measured at the end of a QEC period. The
logical operators are defined as $\bar{X} = \prod_{i\in\Gamma}
\sigma_i^x$, where $\Gamma$ is a path along the center of the
plaquettes crossing horizontally the lattice, and $\bar{Z} =
\prod_{i\in\bar{\Gamma}} \sigma_i^z$, where $\bar{\Gamma}$ is a path
along the edge of the plaquettes crossing the lattice
vertically. Finally, the codewords of the surface code are also easily
written:
\begin{equation}
\left\{| \bar{\uparrow} \rangle = G |F_z\rangle, | \bar{\downarrow}
\rangle = \bar{X} | \bar{\uparrow} \rangle \right\},
\label{eq:GF}
\end{equation}
where $G = \frac{1} {\sqrt{2^{N_{A}}}} \prod_{\diamondsuit} \left( 1 +
A_{\diamondsuit} \right)$, $N_{A}$ is the total number of possible
stars, and $|F_z\rangle$ is the ferromagnetic state along the positive
$z$ direction, namely, $|F_z\rangle = \prod_{i=1}^N
|\uparrow\rangle_{i,z}$.

The Hamiltonian we consider is written as
\begin{equation}
H = H_0 + V,
\end{equation}
where $H_0$ is a free bosonic Hamiltonian, $H_0 = \sum_{\bf k \neq
0}\, \omega_{\bf k} a_{\bf k}^\dagger a_{\bf k} $, and $V =
\frac{\lambda}{2} \sum_{\bf r} f ({\bf r}) \sigma_{\bf r}^x$,
where ${\bf r}$ denotes the spatial location of a qubit and $f$ is a
local bosonic operator,
\begin{equation}
f ({\bf r}) = \frac{\omega_0}{L^D} \sum_{{\bf k}\neq 0} \left| {\bf k}
\right|^s \left( e^{i {\bf k} \cdot {\bf r}} a_{\bf k}^{\dagger} +
e^{-i {\bf k} \cdot {\bf r}} a_{\bf k} \right).
\end{equation}
Here, $D$ is the bath spatial dimension, $\omega_{\bf k} = v|{\bf
  k}|$, and $\omega_0$ is a microscopic coupling that makes $f$
dimensionless (we adopt units such that $\hbar=1$). This error model
leads to a remarkably simple evolution operator in the interaction
picture: For a time interval $\Delta$,
with the environment starting at its ground state, 
we have \cite{NERM10}
\begin{eqnarray}
\label{eq:evolution}
\hat{U}(\Delta) & = & e^{-\frac{\lambda^2}{2} N {\cal G}_{\bf
    rr}(\Delta)}\, e^{-\frac{\lambda^2}{2} \sum_{{\bf r} \neq {\bf s}}
  \Phi_{{\bf r}{\bf s}} (\Delta) \sigma_{\bf r}^x \sigma_{\bf s}^x}
\nonumber \\ & & \times\ :e^{-i\lambda \sum_{\bf r} F_{\bf r} (\Delta)
  \sigma_{\bf r}^x}:,
\end{eqnarray}
where $N$ is the total number of qubits, ${\cal G}_{{\bf r}{\bf s}}
(\Delta) = \langle 0 | F_{\bf r} (\Delta)\, F_{\bf s} (\Delta) | 0
\rangle$, $F_{\bf r} (\Delta) = \int_0^\Delta dt\, f ({\bf r},t)$,
$\Phi_{{\bf r}{\bf s}} (\Delta) = {\cal G}_{{\bf r}{\bf s}} (\Delta) +
\int_0^\Delta dt_1 \int_0^{t_{1}} dt_2 \left[ f ({\bf r},t_1), f ({\bf
    s},t_2) \right] $, and $::$ stands for normal ordering. For Ohmic
baths, the correlation function for ${\bf r}\neq{\bf s}$ takes the
simple form \cite{bathcorrelator}
\begin{equation}
\Phi_{{\bf r}{\bf s}}(\Delta) = \left( \frac{\omega_0}{v} \right)^2
\left\{ \begin{array}{ll} \mbox{arcsh} \left( \frac{v \Delta}{|{\bf r}
    - {\bf s}|} \right) + \frac{i\pi}{2}, & 0 < |{\bf r} - {\bf s}| <
  v\Delta, \\ i\, \mbox{arcsin}\left( \frac{v \Delta}{|{\bf r} - {\bf
      s}|} \right), & 0 < v\Delta < |{\bf r} - {\bf s}|. \end{array}
\right.
\label{ohmic-corr}
\end{equation}
Thus, we can introduce a fictitious inverse temperature $\beta =
\frac{1}{2} \left( \frac{\lambda \omega_0}{v} \right)^2$ and rewrite
the intermediate factor that contains the two-spin interaction in
Eq. (\ref{eq:evolution}) as $e^{-\beta \sum_{{\bf r} \neq {\bf s}}
  J_{{\bf r}{\bf s}}\, \sigma_{\bf r}^x \sigma_{\bf s}^x}$, where
$J_{{\bf r}{\bf s}}$ represents an effective antiferromagnetic
coupling
 \cite{comment1}.

To simplify the notation, we assume that the system is prepared
initially in the logical state $|\bar{\uparrow}\rangle$ and the boson
field initial state is the vacuum
\begin{equation}
|\psi_0 \rangle = (G|F_z \rangle) \otimes |0 \rangle.
\end{equation}
We then let the system evolve under the Hamiltonian $H$ until a time
$\Delta$, when an error correction protocol is performed flawlessly.

{\it The assumptions} --Since we are only allowing for bit-flip
errors, the syndrome outcome for the star stabilizers is trivial. For
the plaquettes, in principle, all possible syndromes should be
considered. However, it is useful to look at the most benign evolution
and assume that all plaquette syndromes return a nonerror. This
nonerror syndrome provides an upper bound to the available
computational time and also substantially simplifies the calculation
since it removes from consideration which recovery operation should be
performed to steer the system back to the computational basis.

In QEC theory, it is standard to focus only on the qubit system's
evolution and disregard any change to the environment's state, even
though the latter is a also a quantum system capable of sustaining
correlations. If no extra step is taken to dissipate those
correlations, the environment will keep a memory of events that
happened between the QEC periods. Keeping track of such bath-induced,
long-time correlations between QEC cycles in a fidelity calculation is
a difficult task even for simple, nontopological logical qubit systems
\cite{Novais2008,NERM10}. Thus, to proceed with the calculation, we
consider an extra step to the QEC protocol. In addition to projecting
the quantum computer wave function back to the logical Hilbert space,
{\it we assume that at the end of the QEC step the environment is
  reset to its ground state}. Hence, from this point on, we are
excluding from the calculation any spatial correlation between QEC
periods, as well as memory and spatial correlations between the time
evolution of bras and kets. Physically, this is equivalent to assuming
that the environment thermalizes with an even larger zero-temperature
bath during the QEC period. When we adopt this extra simplifying
assumption, we can conveniently rewrite the nonerror syndrome
projector as
\begin{equation}
P^\prime = |\psi_0 \rangle \langle \psi_0| + \bar{X} |\psi_0 \rangle
\langle \psi_0| \bar{X}.
\label{eq:Pprime}
\end{equation}
By using Eq.~(\ref{eq:Pprime}), it becomes now straightforward to
write an expression for the logical qubit fidelity just after the
syndrome extraction:
\begin{eqnarray}
{\cal F} & = & \frac{|{\cal A}|} {\sqrt{|{\cal A}|^2 + |{\cal B}|^2}},
\end{eqnarray}
where $ {\cal A} = \langle \psi_0 | \hat{U} (\Delta) |\psi_0
\rangle$ and $ {\cal B} = \langle \psi_0 | \bar{X} \hat{U} (\Delta)
|\psi_0 \rangle$.

{\it Fidelity calculation} --Thus, our task now is reduced to evaluate
${\cal A}$ and ${\cal B}$. By using Eq.~(\ref{eq:evolution}), it is
straightforward to show that
%
\begin{equation}
\label{eq:A1}
{\cal A} 
= \chi \left\langle F_z \left| e^{-\beta {\cal H}} G^2
\right| F_z \right\rangle\\
\end{equation}
and
\begin{equation}
\label{eq:A2}
{\cal B}
= \chi \left\langle F_z \left| \bar{X} e^{-\beta {\cal H}}
G^2 \right|F_z \right\rangle\\
\end{equation}
%
where
\begin{equation}
\label{eq:Heff}
{\cal H}
= \sum_{{\bf r} \neq {\bf s}} J_{{\bf r}{\bf s}}\, \sigma_{\bf
  r}^x \sigma_{\bf s}^x,
\end{equation}
%
and $\chi = e^{-\frac{\lambda^2}{2} N{\cal G}_{\bf rr}
  (\Delta)}$. Notice that although we chose to start with a
microscopic model of the environment to make a connection to physical
implementations, we could as well have started by imposing an
effective two-body interaction between qubits such as that defined by
${\cal H}$.

Clearly, when $\beta \to 0$, we have ${\cal B} \to 0$ and ${\cal F}
\to 1$. A perturbative expansion for small $\beta$ is the standard
route to discuss the error threshold. This ``high-temperature''
expansion will be discussed elsewhere. Here we follow a different
route. To understand the opposite limit $\beta \to \infty$, we need to
rewrite $|F_z\rangle$ in the $x$ basis:
\begin{equation}
\label{eq:F}
|F_z\rangle = \prod_{i=1}^N \left( \frac{\left| \uparrow
  \right\rangle_{i,x} + \left| \downarrow \right\rangle _{i,x}}
{\sqrt{2}} \right).
\end{equation}
This dual representation corresponds to a ``low-temperature''
expansion, and it is suitable for describing the regime where error
correlations are strong. Inserting Eq. (\ref{eq:F}) into (\ref{eq:A1})
and (\ref{eq:A2}), we obtain
\begin{eqnarray}
\label{eq:A1-GPhi-1-1-1}
{\cal A} &=& \frac{\chi}{2^N} \sum_S e^{-\beta E_s} \langle S| G^2
|S\rangle,\\
%
%
\label{eq:A2-GPhi-1-1-1}
{\cal B} &= &\frac{\chi}{2^N} \sum_S e^{-\beta E_s} \langle S| \bar{X}
G^2 |S\rangle,
\end{eqnarray}
respectively, where $|S\rangle$ is an element of the $x$ basis and
$E_s = \langle S| {\cal H} |S \rangle$ is its ``energy''. Notice that
$E_s$ may have an imaginary part.

\begin{figure}[t]
\centering \subfigure{
\includegraphics[width=3.5cm]{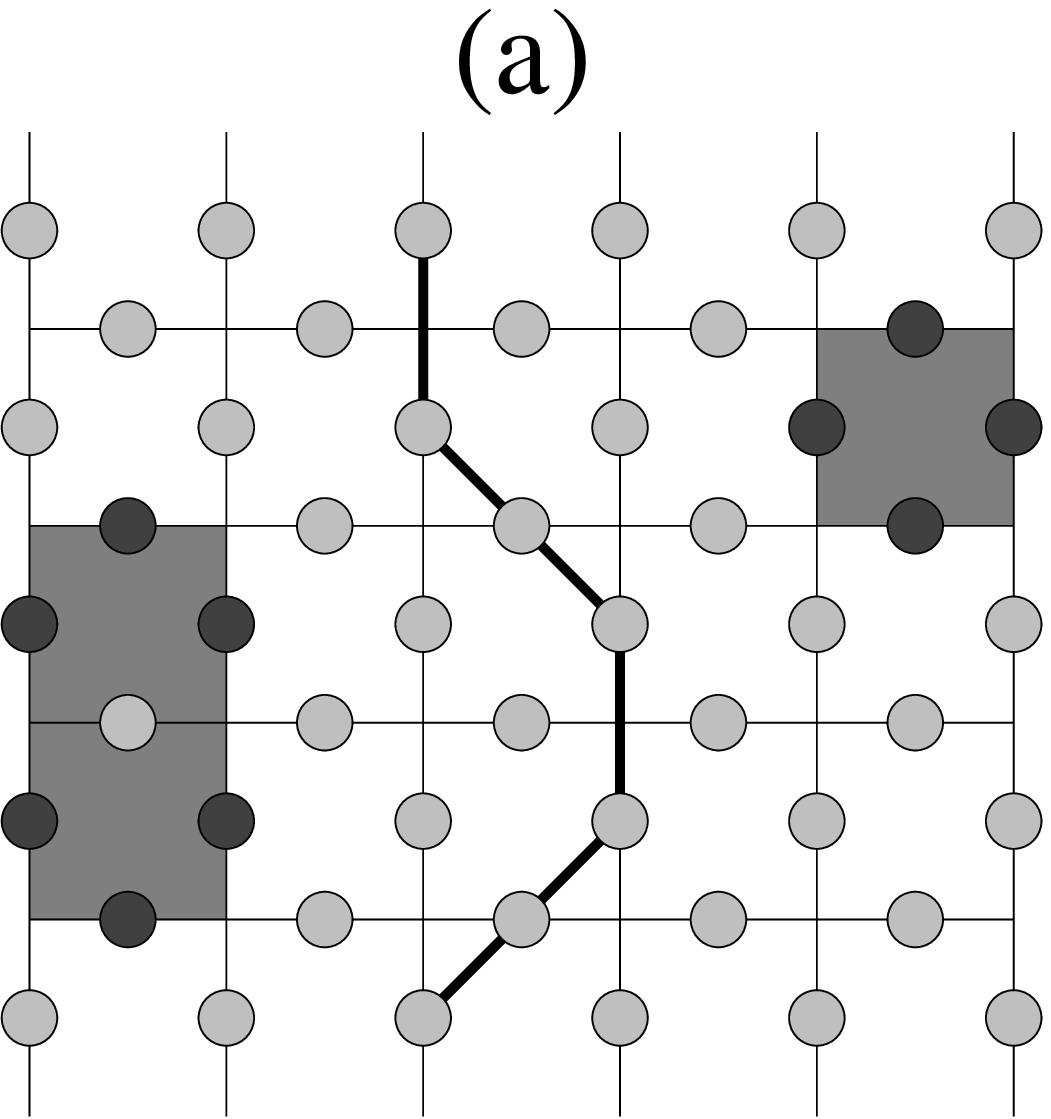}}
\hspace{.5cm}
\subfigure{
\includegraphics[width=3.5cm]{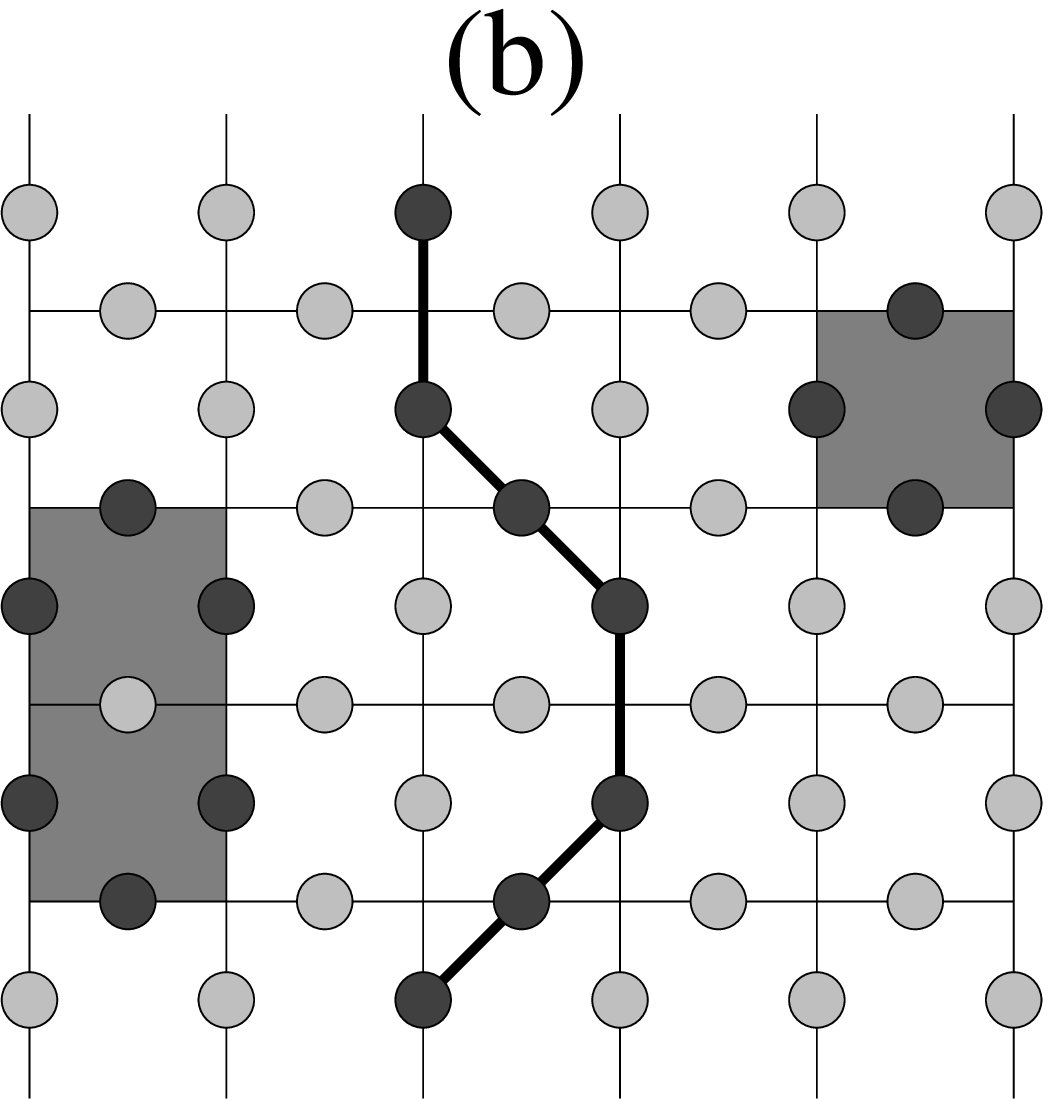}}
\caption{Example of states that contribute to the $S_+$ (a) and $S_-$
  (b) sums [see Eq. (\ref{eq:S12})]. The light (dark) circles indicate
  qubits in $+$ ($-$) $\sigma^x$ eigenstates. The dark line indicates
  the path of the logical error $\bar{Z}_\gamma$, which is active in
  (b) and inactive in (a).}
\label{comparing-energies}
\end{figure}

If we had unrestricted sums in Eqs. (\ref{eq:A1-GPhi-1-1-1}) and
(\ref{eq:A2-GPhi-1-1-1}), we would be essentially discussing a
two-dimensional Ising model. However, $G^2$ projects $|S\rangle$ onto
the subspace of positive stars and, among other things, removes time
reversal state of $\left| S_\pm^\gamma \right\rangle$ from the
sum. This restriction makes the computation of ${\cal A}$ and ${\cal
  B}$ nonstandard. Nevertheless, the action of $\bar{X}$ is to
introduce a sign between two distinct classes of states. To better
understand this, we need a more convenient way to write the states
$|S\rangle$ in the restricted subspace of positive stars. It is not
hard to prove that these states fall into two groups,
$\{|S_+\rangle\}$ and $\{|S_-\rangle\}$, where
\begin{equation}
\label{eq:S12}
|S_+\rangle = \prod_j B_{\square_j} |F_x\rangle \qquad \mbox{and}
\qquad |S_-\rangle = \bar{Z}_\gamma |S_+\rangle. 
\end{equation}
(See Fig. \ref{comparing-energies}.) Here, $\prod_j B_{\square_j}$ is
a product of plaquettes that do not touch a logical error
$\bar{Z}_\gamma$ and $|F_x\rangle$ is the ferromagnetic state in the
$x$ basis. Splitting the terms of the sums in
Eqs. (\ref{eq:A1-GPhi-1-1-1}) and (\ref{eq:A2-GPhi-1-1-1}) between
these two groups of states, we rewrite ${\cal A} = \frac{\chi}{2^N}
({\cal T}_+ + {\cal T}_-)$ and ${\cal B} = \frac{\chi}{2^N} ({\cal
  T}_+ - {\cal T}_-)$, where
\begin{equation}
{\cal T}_{\pm} = \sum_{S_{\pm}} \left\langle S_\pm \left| e^{-\beta {\cal H}}
\right| S_\pm \right\rangle.
\end{equation}
Thus, we have to evaluate the ``free energies'' of the two groups of
states (see Fig.~\ref{comparing-energies}).

The double sum in ${\cal H}$ [see Eq. (\ref{eq:Heff})] runs over
lattice points ${\bf r}$ and ${\bf s}$ inside and outside the path of
the logical operator $\bar{Z}_\gamma$. Thus, let us break these points
into two sets, namely, $\{{\bf r}\} = \{{\bf t}_\gamma\} \oplus \{{\bf
  u}_\gamma\}$ and $\{{\bf s}\} = \{{\bf v}_\gamma\} \oplus \{{\bf
  w}_\gamma\}$, where ${\bf t}_\gamma$ and ${\bf v}_\gamma$ belong to
$\bar{Z}_\gamma$ while ${\bf u}_\gamma$ and ${\bf w}_\gamma$ do
not. Furthermore, let us factor the sum over all $S_+$ into a sum over
paths $\gamma$ and a sum over configurations $S_+^\gamma$ compatible
with a logical operator along this path, namely, $\sum_{S_\pm} =
\sum_\gamma \sum_{S_\pm^\gamma}$. After some simple algebra, we obtain
${\cal T}_{\pm} = \sum_\gamma e^{-\beta
  \epsilon_\gamma}\sum_{S_\pm^\gamma} z^\gamma_{\pm}$, where
$\epsilon_\gamma = \left\langle S_\pm^\gamma \left| \sum_{{\bf t}{\bf
    v}}\, J_{{\bf t}{\bf v}}\, \sigma_{\bf t}^x \sigma_{\bf v}^x
\right| S_\pm^\gamma \right\rangle$ and
\begin{equation}
\label{eq:Zgamma}
z^\gamma_{\pm} = \left\langle S_+^\gamma \left|
e^{-\beta \sum_{{\bf u}\neq {\bf w}} J_{{\bf u}{\bf w}}\, \sigma_{\bf
    u}^x \sigma_{\bf w}^x \mp \beta \sum_{\bf w} h_{\bf w}^\gamma\,
  \sigma_{\bf w}^x} \right| S_+^\gamma \right\rangle,
\end{equation}
with $h_{\bf w}^\gamma = \sum_{\bf t} J_{{\bf t}{\bf w}}$. We can see
that the effect of the logical operator $\bar{Z}_\gamma$ is to
introduce a boundary term represented by the effective local magnetic
field $h_{\bf w}^\gamma$.

We are now in the position to state our definition of the quantum
error threshold. We define the critical parameter $\beta_c$ as the
value of $\beta$ that separates the regime where ${\cal F}=1$ from the
regime where ${\cal F} < 1$ in the {\it thermodynamic} limit
($N\rightarrow \infty$).


{\it Phase transition} --The evaluation of $z^\gamma_{\pm}$ in
Eq.~(\ref{eq:Zgamma}) is a formidable task, and a general answer may
only be achievable through numerical simulations. Hence, we now
restrict our considerations to more manageable effective qubit
interactions. Let us first consider the case of
\begin{equation}
J_{{\bf r}{\bf s}}=\begin{cases} J, & {\bf r}, {\bf s}\ \mbox{nearest
  neighbors},\\ 0 & \mbox{otherwise},
\end{cases}
\label{nn-case}
\end{equation}
where $J$ is real. Such a case is of physical relevance. Any
measurement or operation on the plaquettes and stars can introduce
short-range correlated errors, regardless of the presence of an
environmental bath. For an Ohmic bath, it corresponds to set $v\Delta$
to the order of the lattice spacing.

To show that an order-to-disorder transition indeed exists, let us
consider Eq. (\ref{eq:Zgamma}) with (\ref{nn-case}) in the absence of
the field $h_{\bf w}^\gamma$. Using Eq. (\ref{eq:S12}), we obtain
\begin{equation}
z_{S_\pm^\gamma} = \left\langle F_x \left| \prod_j B_{\square_j}
e^{\beta \sum_{{\bf u}\neq {\bf w}} J_{{\bf u}{\bf w}}\, \sigma_{\bf
    u}^x \sigma_{\bf w}^x} \prod_j B_{\square_j} \right| F_x
\right\rangle,
\end{equation}
where we have multiplied all spins in one of the sublattices by $-1$
to make the model ferromagnetic. The products over plaquettes
introduce a sum over loops where the qubits at these these loops have
negative eigenvalues (they are located at the edges of the shaded
regions in Fig.~\ref{comparing-energies}). Let us label these qubits
by ${\bf a}$ and ${\bf b}$ and use ${\bf c}$ and ${\bf d}$ to label
qubits with positive eigenvalues. Thus
\begin{equation}
\label{eq:z}
z_{S_\pm^\gamma} = e^{\beta \sum_{{\bf a}\neq {\bf b}} J_{{\bf a}{\bf
      b}}}\, e^{\beta \sum_{{\bf c}\neq {\bf d}} J_{{\bf c}{\bf d}}}\,
e^{- \beta \sum_{{\bf a}\neq {\bf c}} J_{{\bf a}{\bf c}}}.
\end{equation}
At low fictitious temperatures, the first two factors in
Eq. (\ref{eq:z}) dominate. The third factor leads to excitations above
the ferromagnetic ground state whose energy is $E_F$. Therefore, we
can go back Eq. (\ref{eq:Zgamma}) and write (for $h_{\bf w}^\gamma=0$)
\begin{equation}
\sum_{S_\pm^\gamma} z^\gamma = e^{\beta E_F} \left( 1 + \sum_{\rm
  loops} e^{-2\beta \sum_{{\bf a}\in {\rm loop}, {\bf c}\notin {\rm
      loop}} J_{{\bf a}{\bf c}}} \right).\label{loop-expansion}
\end{equation}
%

Then, the energy cost of a loop with length $\ell$ is equal to
$4J\ell$. The scaling of the number of loops with $\ell$ is well
known: $N_\ell \sim \mu^\ell$, where $\mu$ is the lattice connectivity
constant (e.g., $\mu \approx 2.638$ for a square lattice)
\cite{connectivity}.
Hence, the contribution of loops with length $\ell$ to
$\sum_{S_\pm^\gamma} z^\gamma$ goes as $\mu^\ell\, e^{-8\beta
  J\ell}$. Clearly, a tradeoff between energy minimization and
``entropy'' maximization takes place, and a phase transition is
expected at $\beta=\beta_c$, where
\begin{equation}
\beta_c \approx \frac{\ln \mu}{8J}. \label{betac}
\end{equation}
For $\beta>\beta_c$ the system is in its ordered phase. Thus, we
expect the system to be sensible to the boundary magnetic field
direction, leading to ${\cal F} < 1$ after the QEC period. In
particular, there is an obvious result to be stated. At the zero
temperature limit $\beta \to \infty$, only the ``ground state''
contributes to the sums (i.e., only the ${\cal T}_{-}$ sum survives).
Hence, $|{\cal A}|=|{\cal B}|$ and ${\cal F}=\frac{1}{2}$.
Conversely, for $\beta<\beta_c$ (therefore $\lambda < \lambda_c$), a
disordered phase develops, leading to the insensibility to the
boundary field and, therefore, ${\cal F}=1$.

There are some roadblocks to studying the general case represented by
the correlator in Eq.~(\ref{ohmic-corr}). The first one is its
imaginary part. It is expected that any imaginary part would have a
negative effect on the fidelity (a conclusion that can be reached by
applying the Schwarz inequality). Even if we disregard this imaginary
part, the magnetic model given by the real part of
Eq.~(\ref{ohmic-corr}) may lead to frustration. However, since the
real part of the correlator is a slow growing function, it is likely
that the model is controlled by boundary effects (as many other
features of topological systems are). Only through future numerical
work will it be possible do address these issues.

In order to gain some insight into the effects of long-range
correlations, we arbitrate a stripped version of the Ohmic model.
Consider the case where qubits on one of the sublattices interact only
with qubits on the other sublattice through the real part of
Eq.~(\ref{ohmic-corr}). This interaction preserves the bipartite
nature of the lattice and we can again map it onto a ferromagnetic
model. We can repeat all steps of the calculation done in the case
with nearest-neighbor interactions. Then, the energy cost of a loop
with length $\ell$ is equal to $(n-2)J\ell$, where $n$ is the number
of sites connected to any lattice site through the interaction (i.e.,
$n$ is related to the range of the interaction). The transition
happens at
\begin{equation}
\beta_c \approx \frac{\ln \mu}{nJ}.
\end{equation}
Note that $\beta_c$ explicitly depends on the interaction range. In a
strict sense, if all qubits in the lattice are in the ``causality
cone'' (and therefore participate in ${\cal H}$), there is never a
threshold. Although this could seem a dismal result, it also shows us
that a finite QEC period introduces an infrared cutoff that creates a
finite transition temperature. The smallest of such transition
temperatures corresponds to the case of nearest neighbors,
Eq.~(\ref{nn-case}). Finally, we note that, under the same assumptions
that we used for the Ohmic bath, super-Ohmic environments are local
and therefore will always yield a finite transition
``temperature''. Conversely, sub-Ohmic baths will require an infrared
cutoff to yield a finite threshold.

In conclusion, in this Letter we studied the fidelity of a logical
qubit encoded with the surface code after one complete QEC period. We
derived a nontrivial mapping to a statistical mechanical problem and
provided an analytical expression for the error threshold for some
correlated models. This mapping provides a promising route for
exploring fault tolerance of topological quantum error correction
codes in the presence of realistic environments.


We thank M. Correa and R. Paszko for helpful discussions and C. Chamon
and R. Raussendorf for insightful comments. E.N. was partially
supported by INCT-IQ and CNPq (Brazil). E.R.M. was supported in part
by the ONR and NSF (USA).



\begin{thebibliography}{99}

\bibitem{DKP2002} E. Dennis, A. Kitaev, A. Landahl, and J. Preskill,
  J.  Math. Phys. (N.Y.) {\bf 43}, 4452 (2002).

\bibitem{DiVincenzo09-a} D. P. DiVincenzo, Phys. Scri. T {\bf 137},
  014020 (2009).

\bibitem{YWC+12} X.-C. Yao {\it et al}., Nature (London) {\bf 482},
  489 (2012).

\bibitem{RH07} R. Raussendorf and J. Harrington, Phys. Rev. Lett. {\bf
  98}, 190504 (2007).

\bibitem{FWH11} A. G. Fowler, D. S. Wang, and L. C. L. Hollenberg,
  Quantum Inf. Comput. {\bf 11}, 8 (2011).

\bibitem{bombin12} H. Bombin, R. S. Andrist, M. Ohzeki, H.
  G. Katzgraber, and M. A. Martin-Delgado, Phys. Rev. X {\bf 2},
  021004 (2012).

\bibitem{Unruh:1995:992} W. G. Unruh, Phys. Rev. A {\bf 51}, 992
  (1995).

\bibitem{Novais:2006:040501} E. Novais and H. U. Baranger,
  Phys. Rev. Lett. {\bf 97}, 040501 (2006).

\bibitem{Novais:2007:040501} E. Novais, E. R. Mucciolo, and
  H. U. Baranger, Phys.  Rev. Lett. {\bf 98}, 040501 (2007).

\bibitem{Novais2008} E. Novais, E. R. Mucciolo, and H. U. Baranger,
  Phys. Rev. A {[\bf 78} 012314 (2008).

\bibitem{Ng:0810.4953v1} H. K. Ng and J. Preskill, Phys. Rev. A {\bf
  79}, 032318 (2009).

\bibitem{NERM10} E. Novais, E. R. Mucciolo, and H. U. Baranger,
  Phys. Rev. A {\bf 82}, 020303(R) (2010).

\bibitem{Preskill.2012} J. Preskill, technical report No. CALT
  68-2881, 2012; arXiv:1207.6131.

\bibitem{Kogut} J. Kogut, Rev. Mod. Phys. {\bf 51}, 659 (1979).

\bibitem{Gottesman} D. Gottesman, Phys. Rev. A {\bf 54}, 1862 (1996);
  A. R. Calderbank, E. M. Rains, P. W. Shor, and N. J. A.  Sloane,
  Phys. Rev. Lett. {\bf 78}, 405 (1997).

\bibitem{bathcorrelator} D. L\'{o}pez, E. R. Mucciolo, and E. Novais
  (in preparation).

\bibitem{comment1} When the initial state of the environment is a
  thermal state, a thermal length $\xi_T = \hbar v/k_BT$, added to the
  bosonic propagator of Eq. (\ref{eq:evolution}), causes correlation
  functions to decay exponentially over distances larger than
  $\xi_T$. Such a situation can be readily included in the effective
  model of Eq. (\ref{eq:Heff})

\bibitem{connectivity} N. Madras and G. Slade, {\it The Self-Avoiding
  Walk} (Birkh\"auser, Boston, 1996).

\end{thebibliography}
\end{document}